# Layered Compounds BaM$_2$Ge$_4$Ch$_6$ (M = Rh, Ir and Ch = S, Se) with Pyrite-Type Building Blocks and Ge-Ch Heteromolecule-Like Anions

*Hechang Lei,[1] Jun-ichi Yamaura,[2] Jiangang Guo,[1] Yanpeng Qi,[1] Yoshitake Toda,[1] and Hideo Hosono,[1,2,3,*]*

[1] Frontier Research Center, Tokyo Institute of Technology, Yokohama 226-8503, Japan

[2] Materials Research Center for Element Strategy, Tokyo Institute of Technology, Yokohama 226-8503, Japan

[3] Materials and Structures Laboratory, Tokyo Institute of Technology, Yokohama 226-8503, Japan

**Abstract**

Structure and chemical features of layered compounds BaM$_2$Ge$_4$Ch$_6$ (M = Rh, Ir; Ch = S, Se) synthesized under high-pressure and high-temperature method are systematically studied. These compounds crystallize in an orthorhombic phase with space group *Pbca* (No. 61). The remarkable structural feature is to have the M-Ge-Ch pyrite-type building units, stacking with Ba-Ch layers alternatively along *c* axis. It is very rare and novel that pyrite-type subunits as the building blocks in the layered compounds. Theoretical calculations and experimental results indicate that there are strongly polarized covalent bonds between Ge and Ch atoms, forming heteromolecule-like anions in these compounds. Moreover, Ge atoms in this structure exhibit unusual valence state (~ +1) due to the tetrahedral coordination environment of Ge atoms along with M and Ch atoms simultaneously.

* Corresponding author.

hosono@msl.titech.ac.jp

**Introduction**

Hundreds of inorganic compounds with $MCh_2$, $MX'_2$, or $MChX'$ formulas (M = post transition metals, Ch = chalcogens, and X' = tetrels or pnictogens) have pyrite-type structure, which originates from $FeS_2$ and is the most common structure of the sulfide minerals. The structure is composed of a three dimensional network of corner-sharing M centered octahedra with Ch and/or X' atoms at the vertices. The most remarkable feature is the bonding state between two Ch (X') atoms, such as S-S bond in $FeS_2$. The valence state and strength of this molecule-like anion (dimer) can be changed with different bond length of dimer and also influenced by the relative position of $d(e_g)$ level of transition metal to the $\pi^*$ orbital of dimer.[1] This kind of variation of anionic dimer has important effects on the physical properties of pyrite-type compounds. For example, in $NiS_{2-x}Se_x$, there is metal-insulator transition with increasing $x$ and the bonding-antibonding splitting in the S-S (Se-Se) dimer is identified as the main parameter controlling the size of the charge gap.[2,3] On the other hand, the Ir filling in pyrite-type $Ir_xSe_2$ probably enables to control the superconductivity because the electronic state is closely related to the distance of Se-Se dimer.[4] Moreover, in $Ir_{0.94-x}Rh_xSe_2$, when Se-Se dimer becomes destabilized (weaken) accompanying with partial electron transfer from the Ir/Rh to the Se, there is an enhanced superconducting $T_c$ in this system.[5]

In many of Ge-Ch compounds, owing to the chemical similarity between Ge and Si as well as O and Ch, a Ge atom tends to be coordinated by four Ch atoms to form Ge-Ch tetrahedron $GeCh_4$, such as $Eu_2GeS_4$,[6] $PdGeS_3$,[7] $Mn_2GeSe_4$,[8] and $La_3CuGeSe_7$.[9] This building block is also an important structural unit in chalcogenide glasses, such as in binary $GeCh_2$,[10] and ternary $GeS_2$-$SbS_3$ as well as $Ga_2Se_3$-$GeSe_2$ systems.[11,12] In the Ge-Ch tetrahedron, Ge usually exhibits positive valence state (+4) in polarized covalent bond between Ge and Ch.

For ternary pyrite-type compounds, the compounds including Ge and Ch atoms simultaneously are scarce, and only two compounds PtGeS and PtGeSe are reported until now.[13] They have cobaltite structure, a ternary variant of the pyrite-type structure.[13] Although in these compounds Ge atoms are located at the center of tetrahedron, similar to other Ge-Ch compounds, the atoms at vertices of tetrahedron are not only Ch but also Pt. In our search of new quaternary Ge-S/Se compounds, a new family of layered compounds $BaM_2Ge_4Ch_6$ (M = Rh, Ir; Ch = S, Se) with pyrite-type building units is discovered. This structure is built up with stacking the M-Ge-Ch pyrite-type slabs and Ba-Ch layers alternatively along $c$ axis. This is the first case that pyrite-type subunits are utilized to build the compounds with complex structure. More importantly, theoretical calculations and experimental results indicate that there is a strong polarized covalent bond (dimer) between Ge and Ch atoms. Moreover, the tetrahedral coordination of Ge atom by both M and Ch atoms results in the unusually small positive valence state of Ge (~ +1) in these compounds, which is significantly different from the valence state of Ge (+4) in other Ge-Ch compounds with only Ge-Ch tetrahedron. This can be ascribed to the different electronegativity of Rh, Ge and Ch atoms and special coordination environment of Ge atom.

**Experimental Section**

**Synthesis.** $BaM_2Ge_4Ch_6$ polycrystallines were synthesized using high-pressure and high-temperature method. First, the BaS/BaSe precursors were prepared by sealing Ba and S/Se into silica tubes and sintering at 1073 K for 15h. The obtained materials were mixed with stoichiometric amounts of Rh/Ir, Ge and S/Se, well ground and then pelletized. The pellet was loaded into a $h$-BN capsule and then heated at 1473 K and 5 GPa for 2 h using a belt-type high-pressure apparatus. All starting materials and precursors for the synthesis were prepared in a glove box filled with purified Ar gas ($H_2O$, $O_2$ < 1ppm). The single crystal of $BaRh_2Ge_4S_6$ with a typical size up to 0.05×0.05×0.01 mm was grown by prolonging the annealing time (12 h) under high pressure.

**X-ray Powder Diffaction and Elemental Analysis.** The powder x-ray diffraction (PXRD) patterns of $BaM_2Ge_4Ch_6$ filling capillary tubes were collected by using a Bruker diffractometer D8 ADVANCE using a Mo-$K\alpha$ radiation (see Supporting Information). Rietveld refinement of the XRD patterns was performed using the code TOPAS4.[14] The chemical compositions of the samples were examined by electron probe microscope analysis (EPMA) with backscattered electron (BSE) mode. Sample analysis confirmed the presence of only Ba, Rh/Ir, Ge and S/Se.

**Structural Determination.** Single-crystal X-ray diffraction experiments were carried out on a curved imaging plate (Rigaku R-AXIS RAPID-II) using a Mo-$K\alpha$ radiation generated by a rotating anode with a confocal mirror (Rigaku VariMax). Data integration and absorption correction were performed with RAPID-AUTO. The charge flipping method was used to decide initial structural parameters.[15] Structural parameters were refined using the program SHELXL based on the full-matrix least-squares method.[16] Anisotropic displacement parameters were applied for all atoms.

**XPS Spectroscopy.** X-ray photoelectron spectroscopy (XPS) measurements were performed using a hemispherical analyzer (Omicron, EA125) with non-monochromatic X-ray (Al $K_\alpha$ line, $h\nu$ = 1486.7 eV, $\Delta E$ = 1.0 eV) sources. Samples were placed in an ultra-high vacuum apparatus with a base pressure of $2\times10^{-8}$ Pa. The energy scale of XPS spectra was calibrated to the Au $4f_{7/2}$ peak at 84.0 eV.

**Theoretical Calculations.** First-principle electronic-structure calculations were performed using experimental crystallographic parameters within the full-potential linearized augmented-plane-wave (FP-LAPW) method implemented in the WIEN2k package.[17,18] The general gradient approximation (GGA) proposed by Perdew et al. was used for exchange-correlation potential.[19] The product of the muffin tin radius ($R_{MT}$) and the largest wave number of the basis set ($K_{max}$) is fixed at 7.0 for all of calculations. We employed the following $R_{MT}$: $R_{MT\text{-}Ba}$ = 2.5 Bohr for all of four compounds, $R_{MT\text{-}Rh}$ = 2.29/2.30 Bohr for $BaRh_2Ge_4S_6$/$BaRh_2Ge_4Se_6$, $R_{MT\text{-}Ir}$ = 2.37/2.32 Bohr for $BaIr_2Ge_4S_6$/$BaIr_2Ge_4Se_6$, $R_{MT\text{-}Ge}$ = 2.18/2.19/2.14/2.10 Bohr for $BaRh_2Ge_4S_6$/$BaRh_2Ge_4Se_6$/$BaIr_2Ge_4S_6$/$BaIr_2Ge_4Se_6$, $R_{MT\text{-}S}$ = 1.92/1.85 Bohr for $BaRh_2Ge_4S_6$/$BaIr_2Ge_4S_6$, and $R_{MT\text{-}Se}$ = 2.19/2.23 Bohr for $BaRh_2Ge_4Se_6$/$BaIr_2Ge_4Se_6$, respectively. Self-consistency was carried out on $8\times8\times12$ k-point meshes in the whole Brillouin zone. The energy convergence was set to be $10^{-4}$ Ry for the self-consistency. Additionally, electronic structure calculations and bonding analysis were carried out for $BaRh_2Ge_4S_6$ using the tight binding−linear muffin tin orbitals−atomic sphere approximation (TB-LMTO-ASA) program package.[20] Radii of the atomic spheres and interstitial empty spheres were determined by the procedures implemented in the TB-LMTO-ASA programs. The k-space integration was performed by the tetrahedron method.[21] The Barth-Hedin exchange potential was employed for LDA calculations.[22] The radial scalar-relativistic Dirac equation was solved to obtain the partial waves. A basis set containing Ba($6s,5d,4f$), Rh($5s,5p,4d$), Ge($4s,4p$) and S($3s,3p$) orbitals was employed for a self-consistent calculation with Ba($6p$), Rh($4f$), Ge($4d$) and S($3d$) functions being down-folded, as automatically selected by the TB-LMTO-ASA program. For bonding analysis, the energy contributions of all electronic states for selected atom pairs were evaluated with a crystal orbital Hamiltonian population (COHP) analysis.[23] Integration up to Fermi level yielded -ICOHP values as measures of relative overlap populations. The electron localization function (ELF)[24-26] was evaluated with modules implemented within the TB-LMTO-ASA program package. The VESTA program was used for visualization of ELF isosurfaces.[27]

**Results and Discussion**

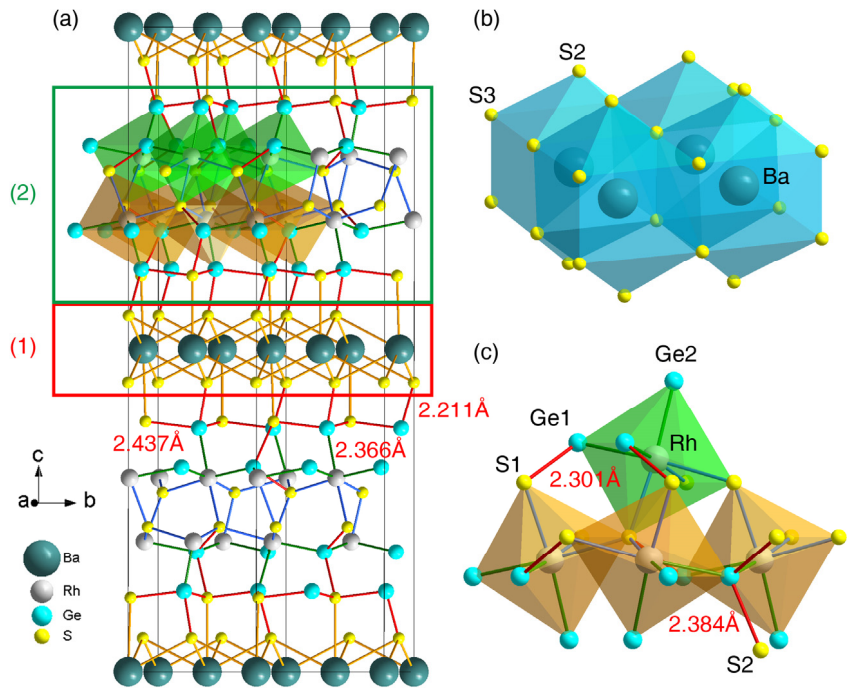

Figure 1. (a) Structure of BaRh$_2$Ge$_4$S$_6$. The octahedron of Rh-Ge/S is emphasized. The atoms are marked as follows: Ba, dark cyan; Rh, white; Ge, light blue; S, yellow. The building blocks (1) and (2) (see text) are emphasized by red and green rectangles. (b) The view of a section of the Ba-S layer. Ba is located in the dodecahedron of S. (c) The view of part of Rh-Ge/S layer. The Ge-S bonds are labeled in red and the corresponding bond lengths are shown.

**Structural Description.** According to single-crystal X-ray diffraction analyses, $BaRh_2Ge_4S_6$ is a layered compound, which can be described well using the orthorhombic space group *Pbca* (No. 61) ($a \approx 5.95$ Å, $b \approx 5.89$ Å, $c \approx 29.20$ Å, and $Z = 4$) with seven independent atom sites (Table 1 and 2). As shown in Figure 1a, this structure contains two building units stacking along $c$ axis alternatively: (1) Ba-S layers and (2) Rh-Ge/S slabs. For the former one, one Ba atom is surrounded by ten S atoms, and the formed Ba-S2/S3 dodecahedra connect each other with face-sharing and expand along $ab$ plane (Figure 1b). The bond lengths between Ba and S2/S3, $d_{Ba-S2/S3}$, spread out in the range between 3.229 and 3.361 Å (Table S1 of the Supporting Information), essentially corresponding to the sum of ionic radii. It indicates that the bond type of Ba-S should be ionic. Moreover, they are slightly larger than those in BaS at ambient ($d_{Ba-S} \approx 3.194$ Å, 6-coordination) and high pressure ($d_{Ba-S} \approx 3.159$ Å, 8-coordination), but comparable to the values in $BaS_3$ ($d_{Ba-S} \approx 3.204$ - 3.541 Å, 12-coordination).[28,29] It could be partially ascribed to the rather high coordination number (10) of Ba in $BaRh_2Ge_4S_6$, which increases the steric effects.

Table 1. Crystal Data and Structure Refinements for $BaRh_2Ge_4S_6$

| | |
|---|---|
| formula | $BaRh_2Ge_4S_6$ |
| formula weight (g/mol) | 826.09 |
| radiation | Mo-$K\alpha$ |
| crystal system | orthorhombic |
| space group | *Pbca* (No. 61) |
| Z | 4 |
| $a$ (Å) | 5.9512(1) |
| $b$ (Å) | 5.8941(1) |
| $c$ (Å) | 29.2011(5) |
| volume (Å$^3$) | 1024.30(3) |
| calculated density (g cm$^{-3}$) | 5.36 |
| absorption coeff. (mm$^{-1}$) | 19.7 |
| no. of reflections collected/$R_{int}$ | 12779/0.020 |
| data/parametes | 3836/62 |
| GOF | 1.034 |
| $R_1/wR_2$ [$I>2\sigma(I)$] | 0.024/0.045 |
| extinction coefficient | 0.0007 |
| largest diff. peak/hole (e Å$^{-3}$) | 1.29/1.82 |

Table 2. Atomic Coordinates and Equivalent Isotropic Displacement Parameters for BaRh$_2$Ge$_4$S$_6$

| atom | Wyckoff | $x$ | $y$ | $z$ | $U_{eq}{}^a$ (×10$^3$ Å$^2$) |
|---|---|---|---|---|---|
| Ba | 4a | 0 | 0 | 0 | 10.33(3) |
| Rh | 8c | 0.00302(2) | 0.50948(2) | 0.200697(5) | 4.07(3) |
| Ge1 | 8c | 0.37941(3) | 0.38733(3) | 0.184156(7) | 5.60(4) |
| Ge2 | 8c | 0.08079(3) | 0.91701(4) | 0.376698(7) | 7.02(4) |
| S1 | 8c | 0.11946(7) | 0.62410(7) | 0.27672(2) | 5.15(7) |
| S2 | 8c | 0.03560(8) | 0.50918(8) | 0.38690(2) | 7.22(7) |
| S3 | 8c | 0.00525(8) | 0.49642(8) | 0.05140(2) | 9.25(8) |

$^a U_{eq}$ is the equivalent isotropic displacement factor defined as one-third of the trace of the orthogonalized $U_{ij}$ tensor

On the other hand, the apparent complexities of this structure arise from the Rh-Ge/S slabs. First, there are two layers of Rh atoms and in each layer Rh atoms form distorted square net (Figure S1 of the Supporting Information). The relative displacement of the second layer to the first layer is about (0, 0.5b, 0.1c). It should be noted that the interlayer distances of Rh atoms are slightly smaller than intralayer ones. One Rh atom is coordinated with three Ge atoms (2×Ge1 and 1×Ge2) and three S atoms (3×S1), forming one Rh-Ge/S octahedron (Figure 1c). In the Rh-Ge/S octahedron, the Ge and S are ordered with facial arrangement. Because of distinctive bond lengths for all Rh-Ge/S bonds, especially between Rh-Ge and Rh-S ones (Table S1 of the Supporting Information), the octahedron of Rh-Ge/S are highly distorted. It is also proved from the scattered bond angles of Ge/S-Rh-Ge/S deviating from the ideal values (90° or 180°) (Table S1 of the Supporting Information). The bond lengths of Rh-S are between 2.422 and 2.475 Å, if assuming that the valence of Rh is +3, the ionic radius of Rh$^{3+}$ is 0.805 Å and the bond lengths between Rh and S is close to the sum of ionic radii (2.505 Å), implying that the chemical bond of Rh-S should be ionic. But the detailed analysis as following indicates that there is strongly polarized covalent bond rather than pure ionic bond existing between Rh and S atoms. The distorted Rh-Ge/S octahedra are connected each other by corner-sharing, forming an extended two-dimensional (2D) network along *ab* plane (Figure 1c).

On the other hand, S1 and Ge1/Ge2 have tetrahedral coordination (Figure 1c). S1 is shared by three Rh-Ge/S octahedra and the fourth bond connects to Ge1. Ge1 bridges two Rh-Ge/S octahedra and connects with two sulfur atoms. One (S1) is in between the Rh layers and another (S2) is out of the Rh-Ge/S slabs. In contrast, Ge2 only connects one Rh-Ge/S octahedron and the rest of three bonds connect with three sulfur atoms (2×S2 and 1×S3). The bond lengths between Ge and S atoms $d_{Ge-S}$ are in the range of 2.211 and 2.437 Å and the bond length of Ge2-S3 is shortest among Ge-S bonds. Providing that the valence of Ba, Rh and S is +2, +3 and -2, respectively, we obtain the valence of Ge is +1. For $S^{2-}$ anion and $Ge^{2+}$ ion, the ionic radius is 1.70 Å and 0.87 Å, respectively, thus the bond lengths between Ge and S are significantly smaller than the sum of ionic radii of $S^{2-}$ and $Ge^{2+}$ (2.57 Å), let alone the sum of ionic radii of $S^{2-}$ and $Ge^{+}$. It clearly indicates that there are strong bonding interactions between Ge and S atoms, suggesting that the bonding type might be not purely ionic but polarized covalent.

**Relation to Pyrite-Type Structure.** At first, layered $BaRh_2Ge_4S_6$-type structure seems an isolated new structure. In fact, it is closely related to the famous pyrite-type structure. Figure 2a shows the structure of pyrite-type compound $RhS_2$. When half of S atoms are replaced by another atoms, such as tetrels or pnictogens, it can form different ordered ternary pyrite-type compounds, known as the cobaltite (CoAsS)-type (Figure 2b) and the ullmannite (NiSbS)-type structures (Figure 2c).[30] PtGeS has the cobaltite-type structure (Figure 2b). If Pt is changed into Rh and half of Rh and S atoms are removed, $RhGe_2S$ subunit will be obtained (Figure 2d). Combining with $BaS_4$ subunit (Figure 2e), finally we can get $BaRh_2Ge_4S_6$ structure (Figure 2f). Therefore, the new synthesized $BaRh_2Ge_4S_6$ may be regarded to have an intergrowth structure of ternary pyrite-type slabs and Ba-S layers along *c* axis. As far as we know, this is the first layered structure containing pyrite-type subunit. It is promising to discover more potential homologous series of layered compounds in this family, like other well-known families such as Ruddlesden-Popper series and Sillén-Aurivillius series.

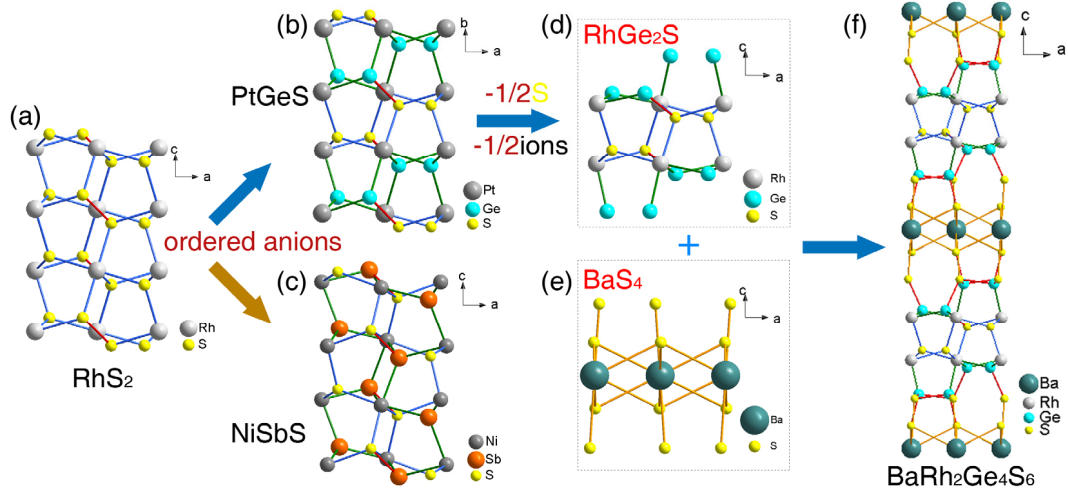

Figure 2. Structures of (a) $RhS_2$, (b) PtGeS, (c) NiSbS, (d) $RhGe_2S$ subunit, (e) $BaS_4$ subunit, and (f) $BaRh_2Ge_4S_6$.

**Electronic Structure Calculations.** In order to understand the physical properties and chemical bonding of $BaRh_2Ge_4S_6$, first-principles and TB-LMTO calculations were carried out. Figure 3 shows the total and partial densities of states (DOS and PDOS) for the $BaRh_2Ge_4S_6$ calculated by the DFT method. The Fermi energy level ($E_F$, set as 0) locates at the edge of valence band, indicating that $BaRh_2Ge_4S_6$ is a semiconductor with the band gap $E_g$ = 1.389 eV. There are three separated sets of bands in the valence region. The lowest region, ranging from -14.5 eV to -11 eV (relative to the $E_F$), is mainly the Ba-5$p$ and S-3$s$ states with a small mixture of Ge-4$s$ state. The second region from -9 eV to -6.6 eV is principally composed of the Ge-4$s$ state with some mixture of S-3$s$ and S-3$p$ states. The third region starting from -5.6 eV and up to $E_F$ is dominated by Rh-4$d$, S-3$p$ and Ge-4$p$ states. Above the $E_F$, the unoccupied states originate mostly from the Rh-3$d$, S-3$p$, Ge-4$s$/4$p$ and Ba-5$d$ states. As shown in Figure 3, the weights of PDOS of Ge and S atoms at the valence band and the conduction band near $E_F$ are similar, suggesting that Ge and S atoms are not pure ionic and should have significantly covalent interactions with neighboring atoms to some extent. It is confirmed by the following bonding analysis.

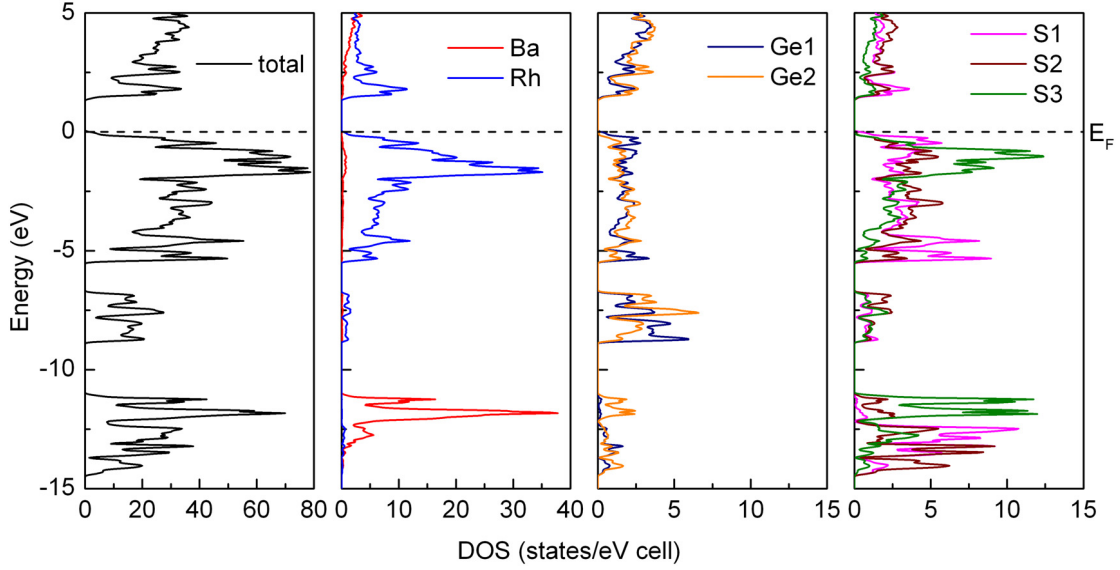

Figure 3. Calculated total DOS and PDOS for $BaRh_2Ge_4S_6$. A horizontal dashed line marks the Fermi level $E_F$, which has been arbitrarily set to zero throughout this work. It should be noted that the ranges are different for the total DOS and PDOS.

**Chemical Bonding.** Because the type and strength of chemical bonds between two Ch (X') elements are crucial in pyrite-type compounds, the chemical bonds in $BaRh_2Ge_4S_6$ especially between Ge and S atoms need to be clarified. In order to obtain a quantitative measure of the bond strength, the crystal orbital Hamiltonian populations (-COHPs) and integrated COHP values (-ICOHPs) up to $E_F$ of select interatomic contacts are calculated (Figure 4 and Table 3). From Figure 4a, it can be seen that the Rh-Ge1(2) bonds are optimized with all bonding states below the Fermi level and anitbonding ones above the Fermi level. But the Rh-S1 bonds are mainly optimized with small antibonding contributions around the Fermi level (Figure 4b). Similar to Rh-Ge/S bonds, Ge-S bonds exhibit mainly bonding states below $E_F$ with slight antibonding states between -8 and -7 eV (Figure 4c and 4d). It should be noted that there is a strong bonding state at about -11 eV for Ge2-S3 contact, which originates from the strong interaction between Ge2-4$s$ and S3-3$s$ states. This strong interaction also leads to the largest -ICOHP (3.83 eV/cell) for Ge2-S3 among all of interatomic contacts (Table 3). Moreover, it can be seen that the values of -ICOHP are roughly inversely proportional to bond lengths, i.e., the larger -ICOHP values correspond to shorter bonds and possible stronger interactions. Besides the strongest Ge2-S3 bonding state, the second highest -ICOHP value is found for the Ge1-S1 bond, indicating the remarkable trend of bonding between Ge and S atoms in the pyrite-type slab. On the other hand, the values of -ICOHP for Rh-Ge1(2) and Rh-S1 are comparable to those ones of Ge-S bonds, indicating that there are also the strong bonding interactions between Rh and Ge/S. In contrast, the small -ICOHP values (≤ 0.35) for the Ba-S2(3) bonding reflect the weak interaction between Ba and S2(3).

Table 3. Selected bond lengths and corresponding -ICOHP for $BaRh_2Ge_4S_6$

| bond | distance (Å) | -ICOHP (eV/cell) | bond | distance (Å) | -ICOHP (eV/cell) |
|---|---|---|---|---|---|
| Ba-S2 | 3.3099(5) | 0.12 | Rh-S1 | 2.4218(5) | 2.05 |
| Ba-S3 | 3.2886(5) | 0.35 | Rh-S1 | 2.4701(4) | 2.12 |
| Ba-S3 | 3.3049(5) | 0.29 | Rh-S1 | 2.4750(5) | 1.89 |
| Ba-S3 | 3.3262(5) | 0.27 | Ge1-S1 | 2.3008(5) | 2.70 |
| Ba-S3 | 3.3607(5) | 0.32 | Ge1-S2 | 2.3844(5) | 2.35 |
| Rh-Ge1 | 2.3839(2) | 2.53 | Ge2-S2 | 2.3655(5) | 2.18 |
| Rh-Ge1 | 2.4019(2) | 2.43 | Ge2-S2 | 2.4371(5) | 1.81 |
| Rh-Ge2 | 2.3777(2) | 2.67 | Ge2-S3 | 2.2113(5) | 3.83 |

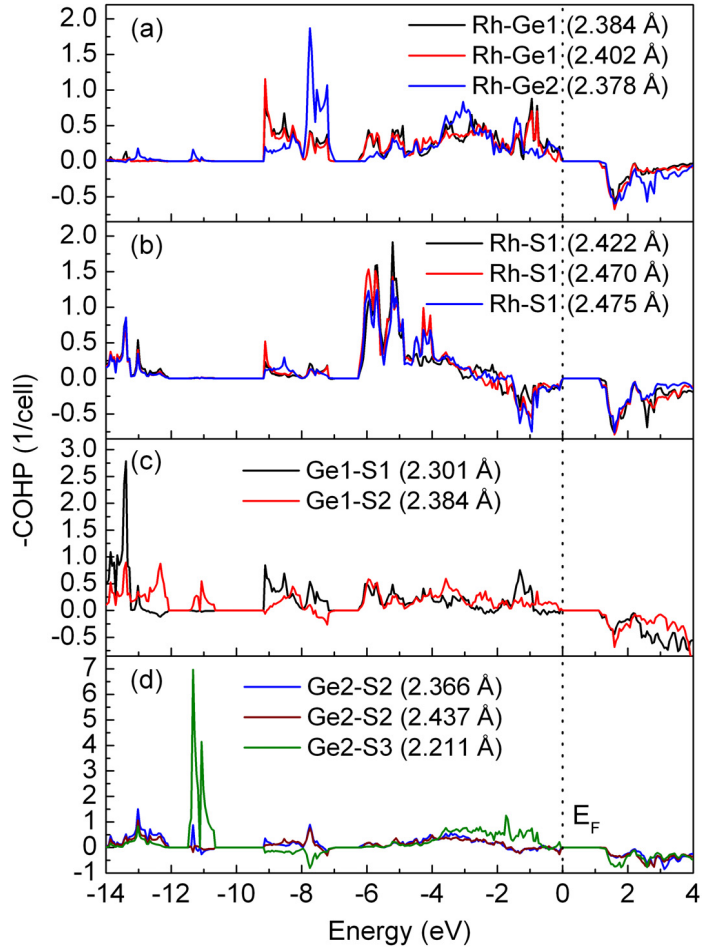

Figure 4. Calculated COHPs of (a) Rh-Ge1(2), (b) Rh-S1, (c) Ge1-S1(2), and (d) Ge2-S2(3) interactions. The distances of each atomic pair are given in brackets. A vertical dashed line marks the Fermi level $E_F$.

Further insight into the nature of the chemical bonding in BaRh$_2$Ge$_4$S$_6$ is provided by the analysis of Electron Localization Function (ELF) as shown in Figure 5. The ELF maxima in the region of the valence electrons indicate either covalent bonds or lone electron pairs. The attractors ① and ② correspond to Rh-Ge1/2 interactions. The ELF maxima are located between Rh-Ge connecting lines, which is typical for the two-center bonds. But they are shifted toward Ge atoms obviously, meaning that the Rh-Ge1/2 bonds are remarkably polarized and there is significant charge transfer from Rh to Ge1/2. This charge transfer suggests that the Allred-Rochow electronegativity scale could be more suitable for this compound than the Pauling electronegativity scale because the Pauling's electronegativity of Ge (2.01) is slightly smaller than that of Rh (2.28) leading to the charge transfer from Ge to Rh[31-33]. On the other hand, the Allred-Rochow's electronegativity of Ge (2.02) is larger than that of Rh (1.45),[34,35] thus it gives the correct polarity of Rh-Ge bonds. The covalent bonding also happens between Rh and S1 atoms (③) with even stronger polarity.

For Ge-S bonding, the localization domains between Ge and S atoms (④-⑦) also shift away from the center of Ge-S connecting lines and are polarized around the S atoms. But this trend is weaker than that in Rh-Ge/S bonds because of smaller difference of electronegativity between Ge (2.02) and S (2.44). Therefore, the type of chemical bonding between Ge1 and S1 atoms in pyrite-type slab is polarized covalent bond. It clearly shows that there are heteromolecule-like anions (dimers) composed by Ge and S atoms, which interact with cation-like Rh atoms. Moreover, because of the unique tetrahedral coordination environment of Ge atoms and the different electronegativity of Rh, Ge and S, it leads to the unusual valence state of Ge in this compound as discussed below.

On the other hand, the lone electron pairs of S2 and S3 atoms near Ba atoms are clearly shown in Figure 5b. However, the shape of lone electron pairs for S3 atom is different from that for S2 because of different coordination environments. The former one has nearly spherical-like shape due to four Ba is surrounding at the bottom part of S3, in contrast, the latter one has ellipsoidal shape. In addition, for S3 atoms, such high density of lone electron pair pushes the valence electron towards Ge2 side, resulting in the weakening of polarity of localization domains and strongest interaction (shortest bond length) between Ge2-S3 contact when compared to bonds (Figure 5b). This could explain the huge peak in the Figure 4d at about -11 eV for Ge2-S3 contact.

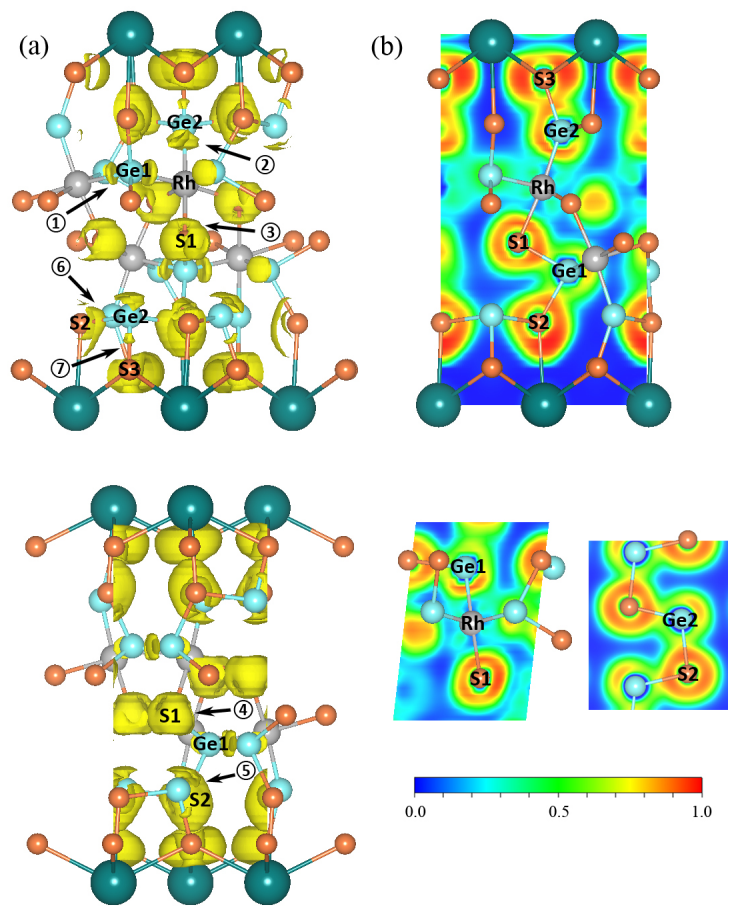

Figure 5. (a) 3D isosurfaces of electron localization function (ELF, η) for BaRh$_2$Ge$_4$S$_6$ (η = 0.8). (b) 2D ELF contour plots for slices of the crystal structure of BaRh$_2$Ge$_4$S$_6$.

**XPS Spectra.** In order to confirm above theoretical analysis, the valence states of Ba, Rh, Ge, and S in BaRh$_2$Ge$_4$S$_6$ were investigated using XPS. Figure 6a shows the XPS spectrum of Ba 3$d_{5/2}$. The binding energy of Ba 3$d_{5/2}$ is 779.5 eV, which is very close to that in BaS (779.8 eV).[36] It indicates that the valence state of Ba should be close to +2. For XPS spectrum of Rh (Figure 7b), there are two peaks at 311.95 and 307.45 eV, corresponding to Rh 3$d_{3/2}$ and 3$d_{5/2}$, respectively. When compared to Rh$_2$O$_3$ (309.1 eV) and Rh$_2$S$_3$ (308.6 eV),[37,38] the peak position of Rh 3$d_{5/2}$ in BaRh$_2$Ge$_4$S$_6$ is shifted to the lower binding energy, implying that the valence state of Rh is smaller than +3. Because the Rh-Ge/S bonds are not pure ionic but polarized covalent and the electronegativity of Ge is smaller than S, both two factors might lead to the smaller valence of Rh than +3. It is consistent with the theoretical analysis. For the XPS spectrum of Ge, the binding energy of Ge 3*d* is about 29.75 eV (Figure 6c), which is significantly smaller than 32.5 eV corresponding to the $Ge^{4+}$ in $GeO_2$.[39] It means that the valence state of Ge is much lower than +4. The binding energy is even smaller than those in GeS (30.5 eV) and GeSe (30.9 eV).[40] On the other hand, this value is larger than that of Ge element (29.0 eV),[40] clearly indicating that the valence of Ge in $BaRh_2Ge_4S_6$ is positive. Schmeisser *et al.* proposed that an average chemical shift per oxidation state for Ge 3*d* core level is 0.85 eV.[41] If taking 29.0 eV as the binding energy for elemental Ge 3*d* core level,[40] the binding energy of $Ge^+$ 3*d* core level should be 29.85 eV, which is very close to the observed value. Thus, it confirms that the valence state of Ge should be close to +1 in $BaRh_2Ge_4S_6$. The small positive valence state of Ge can be understood as follows: Ge atoms are anionic-like in Rh-Ge bonds, thus it should exhibit the negative valence state. But it is compensated by the positive valence state of Ge atoms in Ge-S bonds and finally has a small positive valence state that is still smaller than that in GeS. For the XPS spectrum of S 2*p* (Figure 6d), the spectral peak position (161.45 eV) is similar to that in $TiS_2$ (161.5 eV).[42] It confirms that the valence states of S should be close to -2. On the other hand, the peak position shifts to lower energy when compared to $FeS_2$ (162.4 eV), in which there are unpolarized S-S covalent bonds.[1,43] It is consistent with the theoretical results showing the strong polarity of Ge-S covalent bonds. According to the results of XPS spectra and theoretical calculation, if assuming the valence state of Ba is +2, Rh is +(3-δ), and S is -2, respectively, the charge-balanced formula of $BaRh_2Ge_4S_6$ can be described as $(Ba^{2+})(Rh^{(3-\delta)+})_2(Ge^{+(1+\delta/2)})_4(S^{2-})_6$.

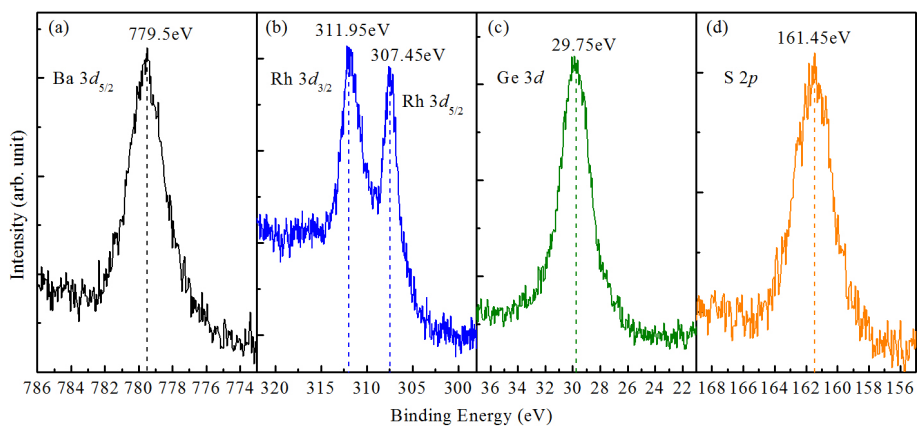

Figure 6. XPS spectra of (a) Ba 3$d$, (b) Rh 3$d$, (c) Ge 3$d$ and (d) S 2$p$ in BaRh$_2$Ge$_4$S$_6$.

**Isostructural Compounds.** In order to extend the phase range of this novel structure type, we also tried to synthesize the (Ba/Sr)(Co/Rh/Ir)$_2$(Ge/Sn)$_4$(S/Se/Te)$_6$ compounds. Among these trials, other three compounds that are isostructural to BaRh$_2$Ge$_4$S$_6$ were discovered: Ba(Rh/Ir)$_2$Ge$_4$(S/Se)$_6$. The PXRD patterns are shown in Figure S2 of the Supporting Information and the fitted lattice parameters are listed in Table 4. The replacement of Rh by Ir does not change the lattice parameters too much, which can be ascribed to the similar ionic/covalent radii between Rh and Ir. On the other hand, the selenides expand the unit cell along all of crystallographic directions because of much larger ionic/covalent radii of Se than those of S. Moreover, for BaIr$_2$Ge$_4$S$_6$, we can not obtain pure phase under current synthesis conditions and secondary phases always exist (Figure S2 of the Supporting Information). These results suggest that BaIr$_2$Ge$_4$S$_6$ may be at the boundary of phase range for this structure.

Table 4. Cell Parameters and reliability factors obtained from powder XRD patterns using Rietveld refinements of BaRh$_2$Ge$_4$S$_6$, BaIr$_2$Ge$_4$S$_6$, BaRh$_2$Ge$_4$Se$_6$, and BaIr$_2$Ge$_4$Se$_6$ (space group *Pbca*)

| formula | BaRh$_2$Ge$_4$S$_6$ | BaRh$_2$Ge$_4$Se$_6$ | BaIr$_2$Ge$_4$S$_6$ | BaIr$_2$Ge$_4$Se$_6$ |
|---|---|---|---|---|
| formula weight (g/mol) | 826.09 | 1107.46 | 1004.71 | 1286.08 |
| a (Å) | 5.9473(2) | 6.1318(3) | 5.9480(4) | 6.1346(4) |
| b (Å) | 5.8891(2) | 6.0700(3) | 5.9148(4) | 6.0870(4) |
| c (Å) | 29.1781(9) | 30.3144(9) | 29.152(2) | 30.279(1) |
| volume (Å$^3$) | 1021.94(6) | 1128.29(8) | 1025.6(1) | 1130.6(1) |
| R$_p$ (%) | 5.74 | 4.32 | 5.43 | 5.08 |
| R$_{wp}$ (%) | 8.25 | 6.41 | 7.85 | 7.54 |
| GOF | 2.79 | 1.68 | 2.03 | 1.80 |

Other three isostructural compounds have similar band structures with different band gaps (Figure S3-S5 of the Supporting Information). The band gap is 1.089, 1.539 and 1.332 eV for BaRh$_2$Ge$_4$Se$_6$, BaIr$_2$Ge$_4$S$_6$, and BaIr$_2$Ge$_4$Se$_6$, respectively. There is a trend that the band gap becomes smaller when S is replaced with Se but increases when Rh is substituted by Ir. This is partially due to the increase of energy dispersion from enhanced orbital overlap when the $p$ states of the chalcogen atoms have contribution to both valence band maximum and conduction band minimum, and the $p$ states become more extended for Se than S.

Although other combinations do not form the isostructural compounds of BaRh$_2$Ge$_4$S$_6$, they form other compounds structurally related to BaRh$_2$Ge$_4$S$_6$. For example, when Ba is replaced by Sr in BaRh$_2$Ge$_4$S$_6$ or Ge is replaced by Sn in BaIr$_2$Ge$_4$Se$_6$, we obtained RhGe$_{1.5}$S$_{1.5}$ or IrSn$_{1.5}$Se$_{1.5}$, an anionic ordered ternary skutterudite compound (Figure 7a).[44,45] They have the same M-Ge(Sn)/S(Se) (M = Rh and Ir) octahedra with facial arrangement of ordered anions and corner-shared connection between octahedra as in some ternary pyrite-type compounds. But the cation M sublattice arranges into distorted simple-cubic geometry, different from the face-centered-cubic (fcc) geometry in pyrite-type compounds. It results in two anionic-anionic dimers for each anion (Figure 7a) in contrast to one dimer in pyrite-type compounds (Figure 7c). On the other hand, when Se is replaced by Te in BaIr$_2$Ge$_4$Se$_6$, IrGeTe is obtained as a main phase, which is a novel phase isostructural to RhGeTe. It has anionic ordered ternary α-NiAs$_2$ (pararammelsbergite) type structure (Figure 7b).[46] M-Ge/Te octahedra with facial arrangement of ordered anions also exist in this structure, similar to pyrite-type compounds. But the half of M sublattice with fcc geometry slides ~ 0.28$b$ along $b$ axis, leading to the edge-shared octahedral connection along sliding plane (Figure 7b). Above results indicate that the basic units of M-Ch(X') octahedron with ordered anions can construct different structures via different arrangement of units. There should be a competing relationship among these phases.

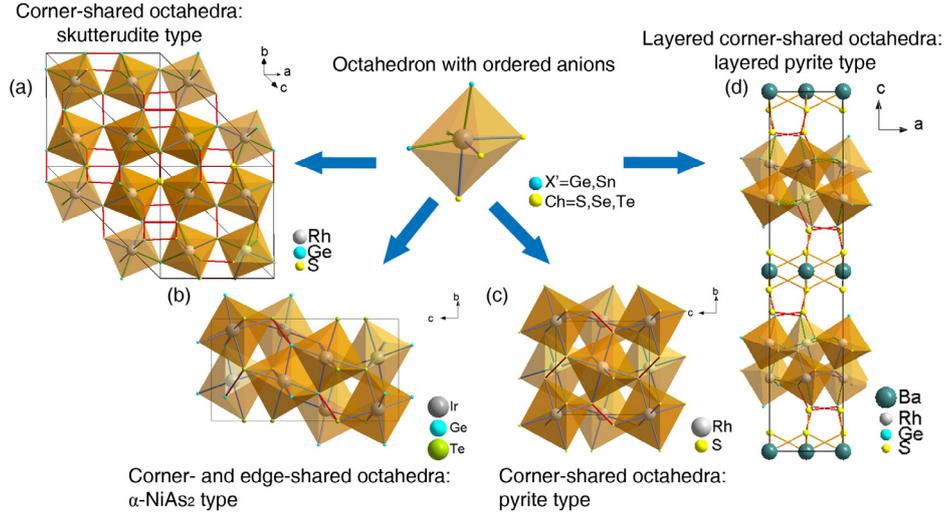

Figure 7. Different structures derived from M-Ch(X') octahedron with ordered anions: (a) skutterudite type $RhGe_{1.5}S_{1.5}$, (b) α-$NiAs_2$ type IrGeTe, (c) pyrite-type $RhS_2$ and (d) layered pyrite-type $BaRh_2Ge_4S_6$.

## Conclusion

In summary, we have discovered series of layered compounds $BaM_2Ge_4Ch_6$ (M = Rh, Ir; Ch = S, Se) which are synthesized by high-pressure and high-temperature method and explored their chemical features systematically. Structural analysis reveals that these compounds contain unprecedented M-Ge-Ch pyrite-type layers. Theoretical calculation indicates that all of them are semiconductors. Experimental results combined with theoretical calculations further suggest that there is strongly polarized covalent bond between Ge and Ch atoms, which have largest electron negativity among Ba, M, Ge and Ch. More interestingly, because of the special tetragonal coordination environment of Ge with M and Ch simultaneously and the different electronegativity of M, Ge and Ch (M < Ge < Ch), Ge exhibits unusual valence state ~ +1. As far as we know, it is the first series of layered compounds with pyrite-type building blocks. It indicates that pyrite-type subunit can be used to build more complex compounds formed under certain non-equilibrium conditions, such as high pressure.

## Acknowledgments

This work was supported by the Funding Program for World-Leading Innovative R&D on Science and Technology (FIRST), Japan, and MEXT Elements Strategy

Initiative to Form Core Research Center.

**Supporting Information**

Supporting Information Available: Listings of X-ray crystallographic file (CIF), selected bond distances and bond angles for $BaRh_2Ge_4S_6$, Powder XRD pattern and fitting results, and calculated total DOS/PDOS for of $BaM_2Ge_4Ch_6$. This material is available free of charge via the Internet at http://pubs.acs.org.

**References**

(1) Jobic, S.; Brec, R.; Rouxel, J. *J. Alloy Compound.* **1992**, *178*, 253-283.

(2) Yao, X.; Honig, J. M.; Hogan, T.; Kannewurf, C.; Spałek, J. *Phys. Rev. B* **1996**, *54*, 17469-17475.

(3) Kuneš, J.; Baldassarre, L.; Schächner, B.; Rabia, K.; Kuntscher, C. A.; Korotin, Dm. M.; Anisimov, V. I.; McLeod, J. A.; Kurmaev, E. Z.; Moewes, A. *Phys. Rev. B* **2010**, *81*, 035122.

(4) Qi, Y. P.; Matsuishi, S.; Guo, J. G.; Mizoguchi, H.; Hosono, H. *Phys. Rev. Lett.* **2012**, *109*, 217002.

(5) Guo, J. G.; Qi, Y. P.; Matsuishi, S.; Hosono, H. *J. Am. Chem. Soc.* **2012**, *134*, 20001-20004.

(6) Tampier, M.; Johrendt, D. *J. Solid State Chem.* **2001**, *158*, 343-348.

(7) Tampier, M.; Johrendt, D. *Chem. Eur. J.* **1998**, *4*, 1829-1833.

(8) Deiseroth, H. J.; Aleksandrov, K. S.; Kremer, R. K. *Z. Anorg. Allg. Chem.* **2005**, *631*, 448-450.

(9) Poduska, K. M.; DiSalvo, F. J.; Min, K.; Halasyamani, P. S. *J. Alloys Compd.* **2002**, *335*, L5-L9.

(10) Elliott, S. R. *Nature* **1991**, *354*, 445-452.

(11) Golovchak, R.; Calvez, L.; Petracovschi, E.; Bureau, B.; Savytskii, D.; Jain, H. *Mater. Chem. Phys.* **2013**, *138*, 909-916.

(12) Lin, C.; Li, Z.; Ying, L.; Xu, Y.; Zhang, P.; Dai, S; Xu, T.; Nie, Q. *J. Phys. Chem. C* **2012**, *116*, 5862-5867.

(13) Entner, P.; Parthé, E. *Acta Crystallogr. B* **1973**, *29*, 1557-1560.

(14) TOPAS Version 4, Bruker AXS: Karlsruhe, Germany, **2007**.

(15) Palatinus, L.; Chapuis, G. *J. Appl. Cryst.* **2007**, *40*, 786-790.

(16) Sheldrick, G. M. *Acta Cryst. A* **2008**, *64*, 112-122.

(17) Blaha, P.; Schwarz, K.; Madsen, G. K. H.; Kvasnicka, D.; Luitz, J. WIEN2k: An Augmented Plane Wave + Local Orbitals Program for Calculating Crystal Properties; Techn. Universität Wien: Wien, Austria, **2001**.

(18) Schwarz, K. *J. Solid State Chem.* **2003**, *176*, 319-328.

(19) Perdew, J. P.; Burke, K.; Ernzerhof, M. *Phys. Rev. Lett.* **1996**, *77*, 3865-3868.

(20) Jepsen, O.; Burkhardt, A.; Andersen, O. K. The Program TB-LMTO-ASA, Version 4.7; Max-Planck-Institut für Festkörperforschung: Stuttgart, Germany, **1999**.

(21) Blöchel, P. E.; Jepsen, O.; Andersen, O. K. *Phys. Rev. B* **1994**, *49*, 16223-16233.

(22) von Barth, U.; Hedin, L. *J. Phys. C: Solid State Phys.* **1972**, *5*, 1629-1642.

(23) Dronskowski, R.; Blöchl, P. E. *J. Phys. Chem.* **1993**, *97*, 8617-8624.

(24) Becke, A. D.; Edgecombe, K. E. *J. Chem. Phys.* **1990**, *92*, 5397-5403.

(25) Savin, A.; Jepsen, O.; Flad, J.; Andersen, O. K.; Preuss, H.; von Schnering, H. G. *Angew. Chem., Int. Ed. Engl.* **1992**, *31*, 187-188.

(26) Savin, A.; Nesper, R.; Wengert, S.; Fassler, T. F. *Angew. Chem., Int. Ed. Engl.* **1997**, *36*, 1808-1832.

(27) Momma, K.; Izumi, F. *J. Appl. Crystallogr.* **2011**, *44*, 1272-1276.

(28) Yamaoka, S.; Shimomura, O.; Nakazawa, H.; Fukunaga, O. *Solid State Commun.* **1980**, *33*, 87-89.

(29) Yamaoka, S.; Lemley, J. T.; Jenks, J. M.; Steinfink, H. *Inorg. Chem.* **1975**, *14*, 129-131.

(30) Foecker, A. J.; Jeitschko, W. *J. Solid State Chem.* **2001**, *162*, 69-78.

(31) Pauling, L. *J. Am. Chem. Soc.* **1932**, *54*, 3570-3582.

(32) Allred, A. L. *J. Inorg. Nucl. Chem.* **1961**, *17*, 215-221.

(33) Dobrotin, R. B. *J. Struct. Chem.* **1963**, *4*, 810-813.

(34) Allred, A. L.; Rochow, E. G. *J. Inorg. Nucl. Chem.* **1958**, *5*, 264-268.

(35) Little, Jr. E. J.; Jones, M. M. *J. Chem. Educ.* **1960**, *37*, 231-233.

(36) Sinharoy, S.; Wolfe, A. L. *J. Electron Spectrosc. Relat. Phenom.* **1980**, *18*, 369-371.

(37) Contour, J. P.; Mouvier, G.; Hoogewijs, M.; Leclere, C. *J. Catal.* **1977**, *48*, 217-228.

(38) Givens, K. E.; Dillard, J. G. *J. Catal.* **1984**, *86*, 108-120.

(39) Barr, T. L.; Mohsenian, M.; Chen, L. M. *Appl. Surf. Sci.* **1991**, *51*, 71-87.

(40) Shalvoy, R. B.; Fisher, G. B.; Stiles, P. J. *Phys. Rev. B* **1977**, *15*, 1680-1697.

(41) Schmeisser, D.; Schnell, R. D.; Bogen, A.; Himpsel, F. J.; Rieger, D.; Landgren, G.; Morar, J. F. *Surf. Sci.* **1986**, *172*, 455-465.

(42) Siriwardane, R. V.; Poston, J. A. *Appl. Surf. Sci.* **1990**, *45*, 131-139.

(43) Volmer-Uebing, M.; Stratmann, M. *Appl. Surf. Sci.* **1992**, *55*, 19-35.

(44) Vaqueiro, P.; Sobany, G. G.; Stindl, M. *J. Solid State Chem.* **2008**, *181*, 768-776.

(45) Fleurial, J.-P.; Caillat, T.; Borshchevsky, A. *Proc. XVI Int. Conf. Thermoelec., IEEE,* Dresden, Germany, **1997**.

(46) Vaqueiro, P.; Sobany, G. G.; Guinet, F.; Leyva-Bailen, P. *Solid State Sci.* **2009**, *11*, 1077-1082.

**Layered Compounds BaM$_2$Ge$_4$Ch$_6$ (M = Rh, Ir and Ch = S, Se) with Pyrite-Type Building Blocks and Ge-Ch Heteromolecule-Like Anions**

*Hechang Lei,[1] Jun-ichi Yamaura,[2] Jiangang Guo,[1] Yanpeng Qi,[1] Yoshitake Toda,[1] and Hideo Hosono,[1,2,3,*]*

[1] Frontier Research Center, Tokyo Institute of Technology, Yokohama 226-8503, Japan

[2] Materials Research Center for Element Strategy, Tokyo Institute of Technology, Yokohama 226-8503, Japan

[3] Materials and Structures Laboratory, Tokyo Institute of Technology, Yokohama 226-8503, Japan

# SUPPORTING INFORMATION

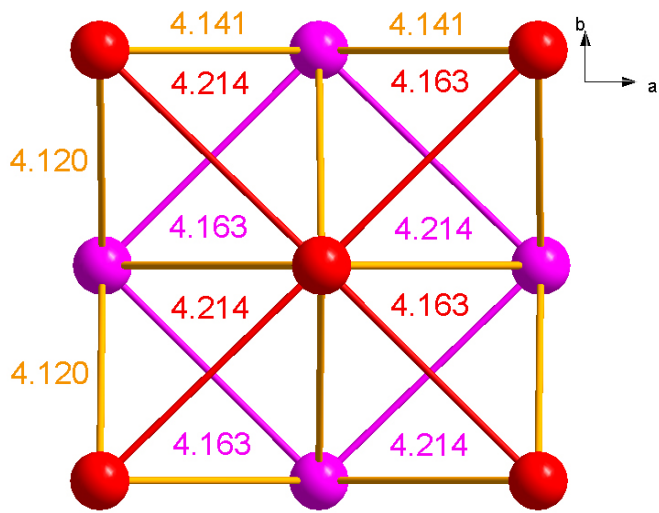

Figure S1. The bond lengths of double Rh layers in BaRh$_2$Ge$_4$S$_6$. The Rh atoms in different layers are labeled in different colors.

Table S1. Selected bond lengths (Å) and angles (°) in BaRh$_2$Ge$_4$S$_6$

| | | | |
|---|---|---|---|
| Ba-S2×2 | 3.3099(5) | Rh-S1 | 2.4218(5) |
| Ba-S3×2 | 3.2886(5) | Rh-S1 | 2.4701(4) |
| Ba-S3×2 | 3.3049(5) | Rh-S1 | 2.4750(5) |
| Ba-S3×2 | 3.3262(5) | Ge1-S1 | 2.3008(5) |
| Ba-S3×2 | 3.3607(5) | Ge1-S2 | 2.3844(5) |
| Rh-Ge1 | 2.3839(2) | Ge2-S2 | 2.3655(5) |
| Rh-Ge1 | 2.4019(2) | Ge2-S2 | 2.4371(5) |
| Rh-Ge2 | 2.3777(2) | Ge2-S3 | 2.2113(5) |
| | | | |
| Ge1-Rh-Ge2 | 86.318(8) | Ge2-Rh-S1 | 88.916(12) |
| Ge1-Rh-Ge1 | 88.027(7) | Ge2-Rh-S1 | 97.041(12) |
| Ge1-Rh-Ge2 | 94.768(9) | Ge2-Rh-S1 | 174.428(13) |
| Ge1-Rh-S1 | 80.859(12) | S1-Rh-S1 | 86.756(15) |
| Ge1-Rh-S1 | 90.064(12) | S1-Rh-S1 | 95.513(15) |
| Ge1-Rh-S1 | 93.046(11) | S1-Rh-S1 | 84.706(14) |
| Ge1-Rh-S1 | 94.003(11) | | |
| Ge1-Rh-S1 | 175.909(12) | | |
| Ge1-Rh-S1 | 176.224(12) | | |

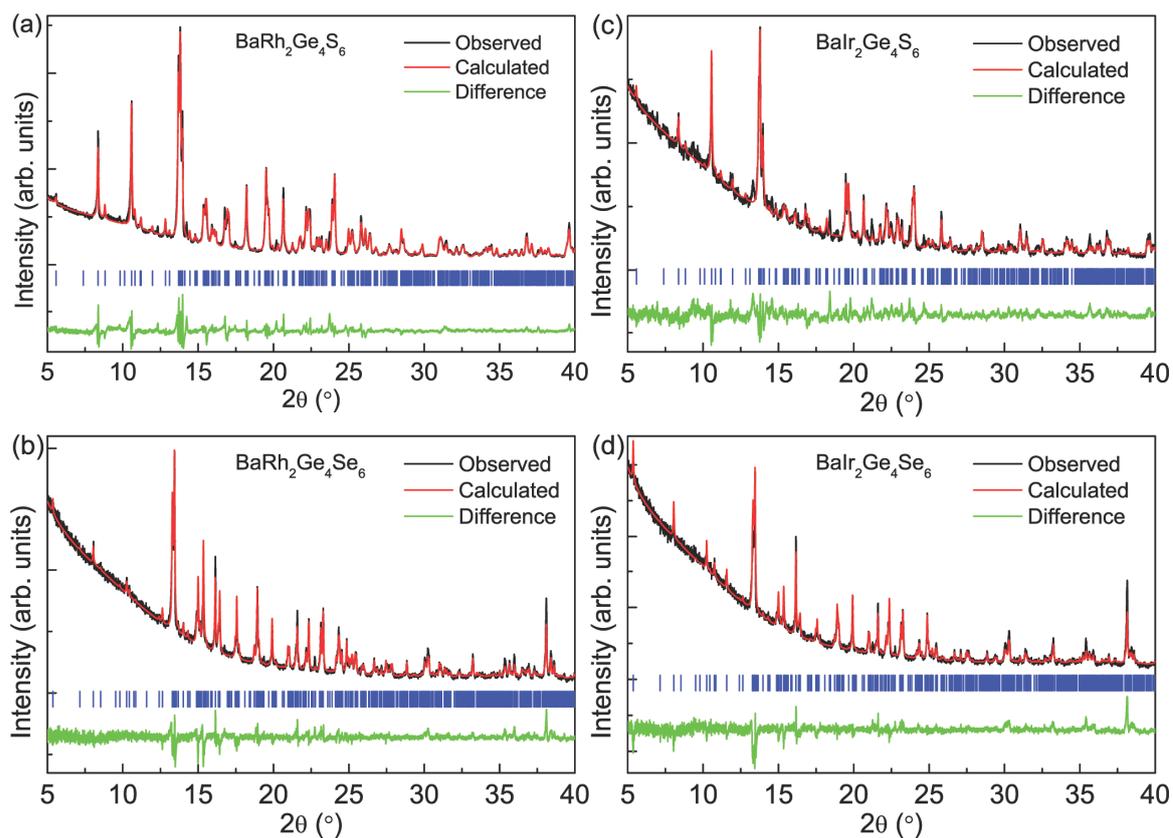

Figure S2. Powder XRD pattern and fitting results of (a) BaRh$_2$Ge$_4$S$_6$, (b) BaRh$_2$Ge$_4$Se$_6$, (c) BaIr$_2$Ge$_4$S$_6$, and (d) BaIr$_2$Ge$_4$Se$_6$.

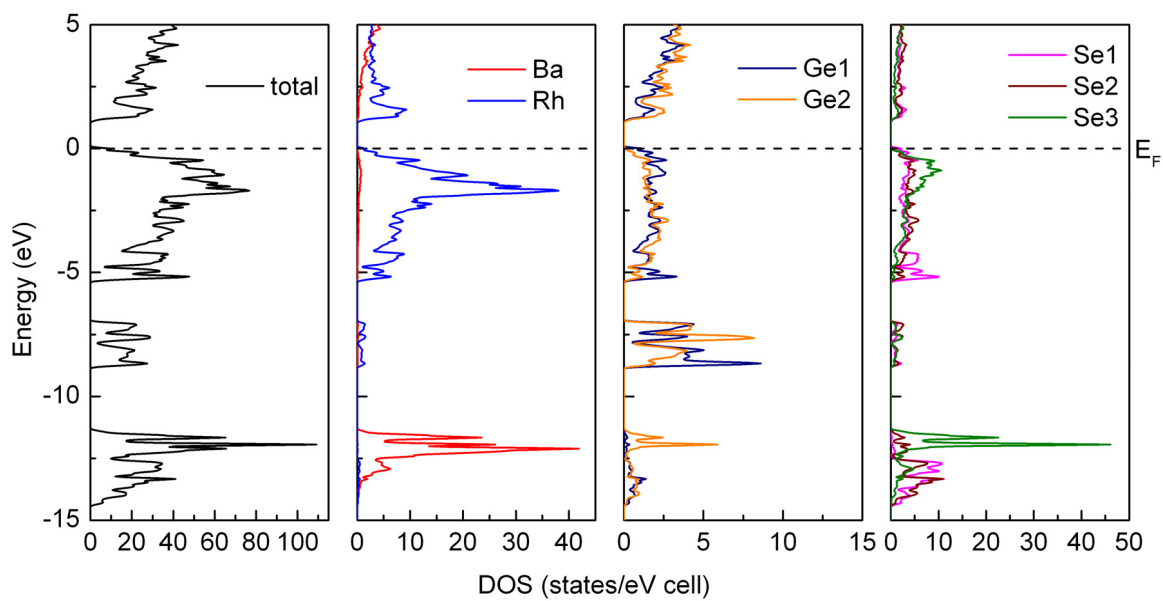

Figure S3. Calculated total DOS and PDOS for BaRh$_2$Ge$_4$Se$_6$. A horizontal dashed line marks the Fermi level $E_F$.

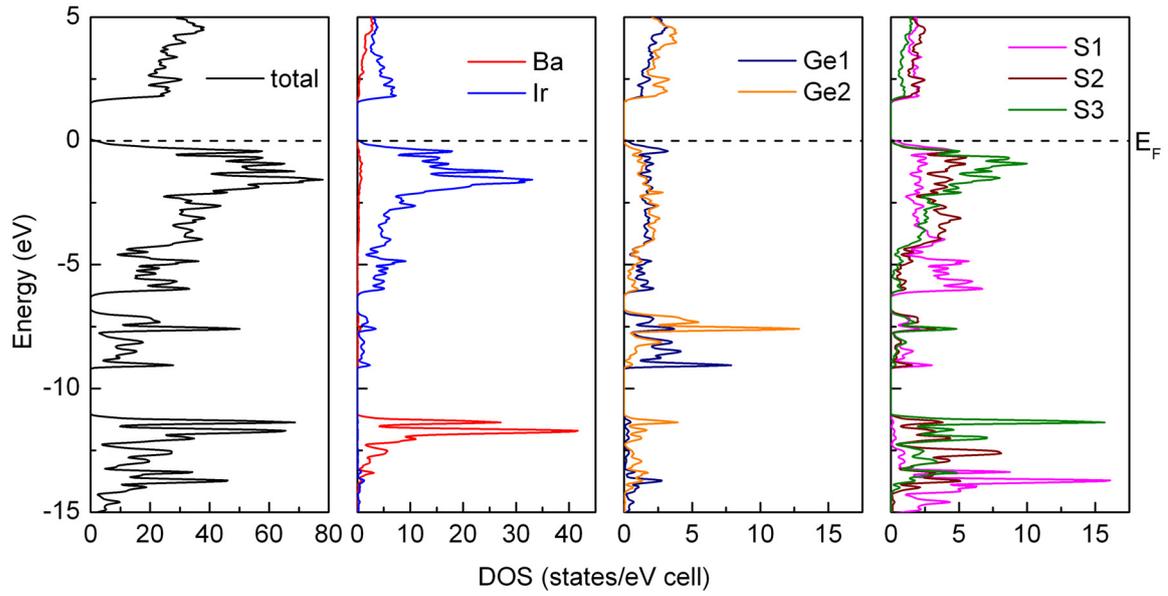

Figure S4. Calculated total DOS and PDOS for BaIr$_2$Ge$_4$S$_6$. A horizontal dashed line marks the Fermi level $E_F$.

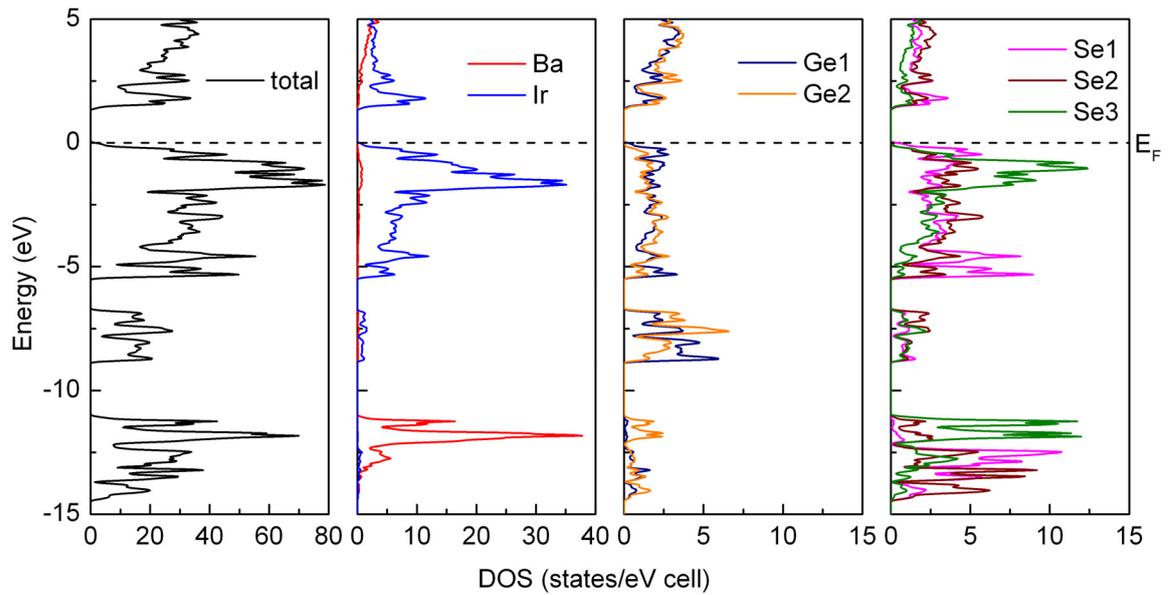

Figure S5. Calculated total DOS and PDOS for BaIr$_2$Ge$_4$Se$_6$. A horizontal dashed line marks the Fermi level $E_F$.